\documentclass[12pt,epsf,showkeys]{article}

\usepackage{graphicx,float}
\usepackage{mathrsfs,array,multirow}
\usepackage{amstext}
\usepackage{subfigure}
\usepackage{bm,latexsym,amsmath,amsfonts,amssymb}
\usepackage[usenames,dvipsnames]{color}
\usepackage{color}
\usepackage[colorlinks=true,linkcolor=blue]{hyperref}
\usepackage{soul}
\usepackage{epsfig}

\newcommand{\be}{\begin{equation}}

\newcommand{\ee}{\end{equation}}

\newcommand{\ba}{\begin{array}}

\newcommand{\ea}{\end{array}}

\begin{document}
\begin{titlepage}
\vspace{.5in}
\begin{flushright}
\end{flushright}
\vspace{0.5cm}

\begin{center}
{\Large\bf Rotating black holes with an anisotropic matter field}\\
\vspace{.4in}

  {$\mbox{Hyeong-Chan\,\, Kim}^{\P}$}\footnote{\it email: hckim@ut.ac.kr},\,\,
  {$\mbox{Bum-Hoon \,\, Lee}^{\S\dag}$}\footnote{\it email: bhl@sogang.ac.kr},\,\,
  {$\mbox{Wonwoo \,\, Lee}^{\S}$}\footnote{\it email: warrior@sogang.ac.kr},\, \,
  {$\mbox{Youngone \,\, Lee}^{\P}$}\footnote{\it email: youngone@ut.ac.kr}\\

\vspace{.3in}

{\small \S \it Center for Quantum Spacetime, Sogang University, Seoul 04107, Korea}\\
{\small \dag \it Department of Physics, Sogang University, Seoul 04107, Korea}\\
{\small \P \it School of Liberal Arts and Sciences, Korea National University of Transportation, Chungju 27469, Korea}\\

\vspace{.5in}
\end{center}
\begin{center}
{\large\bf Abstract}
\end{center}
\begin{center}
\begin{minipage}{4.75in}

{\small \,\,\,\,We present a family of new rotating black hole solutions to Einstein's equations that generalizes
the Kerr-Newman spacetime to include an anisotropic matter.
The geometry is obtained by employing the Newman-Janis algorithm.
In addition to the mass, the charge and the angular momentum, an additional hair exists thanks to the negative
radial pressure of the anisotropic matter.
The properties of the black hole are analyzed in detail including thermodynamics.
This black hole can be used as a better engine than the Kerr-Newman one
in extracting energy.
 }
\end{minipage}
\end{center}
\end{titlepage}

\newpage

\section{Introduction \label{sec1}}

\quad Observations of astrophysical objects show that most of them are rotating.
For this reason, the spacetime geometry describing these rotating objects has attracted attention of researchers for decades~\cite{Kerr:1963ud,Newman:1965my,Gibbons:2004js,Bambi:2013ufa,Herdeiro:2014goa,Adamo:2014baa,Hernandez-Pastora:2017fmg}.
Even if there is no Birkhoff-like theorem for rotating spacetimes, it is believed that the gravitational collapse of a supermassive star forms a rotating black hole eventually.
The energy extraction mechanism from black holes~\cite{Penrose:1969pc, Christodoulou:1970wf, Christodoulou:1972kt, Blandford:1977ds} is a promising candidate describing astrophysical events such as active galactic nuclei, gamma-ray bursts and ultrahigh-energy cosmic rays.

In addition to this aspect, observationally, black hole has recently gained most attention among astrophysical objects, thanks to the observational reports on the shadow of the black hole by the Event Horizon Telescope~\cite{Akiyama:2019cqa, Akiyama:2019bqs, Akiyama:2019fyp}.
The physical origin and the mechanism for explaining the observed image have been widely studied.
Recent detections~\cite{TheLIGOScientific:2016src, Abbott:2016nmj, Abbott:2017vtc} of the gravitational waves coming from binary black hole collisions have also open a new horizon on the studies of astrophysical phenomena and the gravitational theory itself.
The outcome of the collision must be a rotating black hole.
Furthermore, actual astrophysical black holes reside in the background of matters or fields.
Therefore, we need to find a way of describing a realistic black hole that coexists with matter field.

In general, to obtain such a black hole solution, one must solve the Einstein equations analytically or numerically.
However, this task is not easy in an orthodox way even if one assumes a rotating ansatz for metric and matter.
For this reason, we tackle the problem in two steps.
First, we construct a rotating geometry and then obtain matter field satisfying the Einstein equations.
For the first step, we employ the Newman-Janis (NJ) algorithm~\cite{Newman:1965tw, Drake:1998gf,Azreg-Ainou:2014pra}, which has become popular to study the rotational geometry~\cite{Toshmatov:2015npp, Xu:2016jod, Kumar:2017qws,Sakti:2019krw, Shaikh:2019fpu, Contreras:2019nih}.
This algorithm generates a rotating geometry from a known static one.
For the second step, we write the Einstein tensor based on the canonical orthonormal tetrad introduced in Ref.~\cite{Carter:1968ks} and determine the observable quantities, such as energy density and pressure.

In the studies of an astrophysical object, spherical symmetry and isotropy of pressure provide a simple and ideal model.
Thus analyses have been performed mainly for those consisting of a perfect fluid~\cite{Stephani:2003tm,Delgaty:1998uy,Semiz:2008ny}.
However, some matters having locally anisotropic pressure are also allowed in spherically symmetric configuration.
Thereby, studies on anisotropic matter have recently drawn interests~\cite{Ruderman:1972aj,Herrera:1982,Herrera:1997plx,Bowers:1974tgi,Matese:1980zz,Mak:2001eb,Thirukkanesh:2008xc,Ivanov:2002xf,Varela:2010mf,Bekenstein:1971ej,
Cho:2017nhx}.
Thermodynamic properties on anisotropic matter have recently been studied~\cite{Kim:2019ygw,Kim:2019elt}.
Although it is yet to be only theoretically plausible, wormholes consist of anisotropic matter in principle~\cite{Morris:1988cz,Morris:1988tu,Kim:2001ri,Raychaudhuri:1953yv,Kim:2019ojs}.
The studies on the geometrical and the thermodynamical properties of astrophysical objects consisting of anisotropic matter
are becoming important.
In this paper, we consider locally anisotropic fluid as an effective description of matter of which radial pressure is not equal to its angular pressure based on general relativity.

The simplest well-known solution having anisotropic matter field is the Reissner-Nordstr\"{o}m black hole, in which the equation of state satisfies $p_r=-\varepsilon$ in the radial direction and $p_{\theta}=p_{\phi}=\varepsilon$ in the angular directions in a local reference frame.
Generalizing the Reissner-Nordstr\"{o}m solution, a static spherically symmetric black hole solution was found~\cite{Cho:2017nhx}, in which the matter satisfies the anisotropic equation of state.
In this work, we study a rotating generalization of the black hole.

The paper is organized as follows: In Sec.~\ref{sec2}, we present the exact spacetime geometry describing rotating black holes, which generalizes the Kerr-Newman solution.
In Secs.~\ref{sec3} and \ref{sec4}, we analyze the geometric and the thermodynamic properties of the black hole, respectively.
In Sec.~\ref{sec5}, we examine the efficiency of the energy extraction procedure from the black hole.
We summarize and discuss our results in the last section.

\section{A rotating black hole \label{sec2}}

\quad We consider the action
\begin{equation}
I=  \int_{\mathcal M} \sqrt{-g} d^4 x \left[\frac{1}{16\pi}(R-F_{\mu\nu}F^{\mu\nu}) +{\cal L}_m \right] + I_{b}  \,,  \label{action}
\end{equation}
where $R$ is the Ricci scalar of the spacetime ${\mathcal M}$, $F_{\mu\nu}$ is the Maxwell's electromagnetic field tensor, $G=1$ is for simplicity, and $I_{b}$ is the boundary term~\cite{Gibbons:1976ue, Hawking:1995ap}. ${\cal L}_m$ describes effective anisotropic matter fields, which may describe an extra $U(1)$ field as well as diverse dark matters.
Varying Eq.~\eqref{action}, we obtain the Einstein equations
\begin{equation}
G_{\mu\nu}=R_{\mu\nu}-\frac{1}{2}R g_{\mu\nu}=8\pi T_{\mu\nu} \,, \label{einsteineq}
\end{equation}
where $T_{\mu\nu}=\frac{1}{4\pi}(F_{\mu\alpha}F_{\nu}^{\alpha}-\frac{1}{4}g_{\mu\nu}F_{\alpha\beta}F^{\alpha\beta})-2\frac{{\partial \cal L}_m }{\partial g^{\mu\nu}}+ {\cal L}_mg_{\mu\nu}$, and the Maxwell equations
\begin{equation}
\nabla_{\mu}F^{\mu\nu} = \frac{1}{\sqrt{-g}}[\partial_{\mu}(\sqrt{-g}F^{\mu\nu})] =0 \,. \label{maxwelleq}
\end{equation}

The static, spherically symmetric black hole solution with an anisotropic matter field was obtained in \cite{Cho:2017nhx, Kiselev:2002dx},
\begin{equation}
ds^2=  - f(r) dt^2 +  f(r)^{-1} dr^2 + r^2 (d\theta^2 + \sin^2\theta d\psi^2)  \,, \label{}
\end{equation}
where $f(r)=1-\frac{2M}{r} +\frac{Q^2}{r^2} - \frac{K}{r^{2w}}$.
Here, $M$ and $Q$ represent the ADM mass and the total charge of the black hole, respectively, and $K$ is a constant.
The energy-momentum tensor for the anisotropic matter field is diagonal, $T^{\nu}_{\mu}={\rm diag}(-\varepsilon, p_r, p_{\theta}, p_{\psi})$, in which $p_r(r)=-\varepsilon(r)$ and $p_{\theta}(r)=p_{\psi}(r)=w\varepsilon(r)$.
The negative radial pressure allows the anisotropic matter to distribute throughout the entire space from the horizon to infinity.
Therefore, the black hole can be in static configuration with the anisotropic matter field.
The energy density is given by
\begin{equation}
\varepsilon(r) =\frac{Q^2}{8\pi r^{4}}+ \frac{r^{2w}_o}{8\pi r^{2w+2}} \,, \label{}
\end{equation}
where $r_o$ is a charge-like quantity of dimension of length and defined by $r^{2w}_o=(1-2w)K$.
There was an additional analysis for $w=1/2$ case in \cite{Cho:2017nhx}.

Let us begin with the advanced Eddington-Finkelstein coordinates
\begin{equation}
ds^2= - f(r) dv^2 +2dvdr + r^2 (d\theta^2 + \sin^2\theta d\psi^2)  \,. \label{st-EF}
\end{equation}
We now take the null tetrad~\cite{Newman:1961qr} consisting of two real null vectors, $l^{\mu}$ and $n^{\mu}$, and two complex null vectors, $m^{\mu}$ and $\bar{m}^{\mu}$. Here, $m^{\mu}$ and $\bar{m}^{\mu}$ are complex conjugates to each other and satisfy  $l^{\mu}l_{\mu}=n^{\mu}n_{\mu}=0$, $l^{\mu}n_{\mu}=-1$, $m^{\mu}m_{\mu}=0$, $m^{\mu}\bar{m}_{\mu}=1$.
The metric tensor can be expressed in terms of the null tetrad to be
\begin{equation}
g^{\mu\nu}= -l^{\mu}n^{\nu} -n^{\mu}l^{\nu} + m^{\mu}\bar{m}^{\nu} +\bar{m}^{\mu}m^{\nu} \,. \label{metricstandecom}
\end{equation}
In other words, we can directly read off the components of $l_{\mu}$, $n_{\mu}$, $m_{\mu}$ and $\bar{m}_{\mu}$ from the metric~\eqref{st-EF},
\begin{eqnarray}
&& l^{\mu}=\delta^{\mu}_1,~~m^{\mu}=\frac{1}{\sqrt{2} r} \left(\delta^{\mu}_2 -\frac{i}{\sin\theta}\delta^{\mu}_3\right)  \,,  \nonumber \\
&& n^{\mu}=-\delta^{\mu}_0-\frac{1}{2}f(r) \delta^{\mu}_1,~~\bar{m}^{\mu}=\frac{1}{\sqrt{2} r} \left(\delta^{\mu}_2 +\frac{i}{\sin\theta}\delta^{\mu}_3\right) \,.
\end{eqnarray}

In what follows, we explore the geometry of the rotating black hole by using the NJ algorithm~\cite{Newman:1965tw}.
We formally perform complex coordinate transformations,
\begin{equation}
v \rightarrow v'= v+ia\cos\theta \,, \quad  r \rightarrow r' = r +ia\cos\theta \,,
\end{equation}
and then make changes:
\begin{eqnarray}
\frac{2M}{r} \rightarrow \frac{M}{r} + \frac{M}{\bar{r}} = \frac{2Mr'}{r'^2 + a^2 \cos^2\theta} \,, \quad r^2 \rightarrow r\bar{r} =  r'^2 + a^2 \cos^2\theta \,,
\end{eqnarray}
where $a$ is a rotation parameter, the angular momentum per mass.
Then, we get
\begin{eqnarray}
f(r) &\Rightarrow& F(r', \theta); \nonumber \\
1-\frac{2M}{r} +\frac{Q^2}{r^2}- \frac{K}{r^{2w}} &\Rightarrow&
 1 -  \frac{2Mr'-Q^2}{r'^2 + a^2 \cos^2\theta} -\frac{K r'^{2(1-w)}}{(r'^2 +a^2\cos^2\theta)} \,.
\end{eqnarray}
Later in this work, we omit the prime. Then, the null vectors become
\begin{eqnarray}
&& l^{\mu}=\delta^{\mu}_1 \,, \quad n^{\mu}=-\delta^{\mu}_0-\frac{1}{2}F(r,\theta)\delta^{\mu}_1 \,, \nonumber \\
&& m^{\mu}=\frac{\left(-ia\sin\theta(\delta^{\mu}_0+\delta^{\mu}_1)+\delta^{\mu}_2 -\frac{i}{\sin\theta}\delta^{\mu}_3\right)}{(r+i a\cos\theta)\sqrt{2}}  \,, \nonumber \\
&& \bar{m}^{\mu}=\frac{\left(ia\sin\theta(\delta^{\mu}_0+\delta^{\mu}_1)+\delta^{\mu}_2 +\frac{i}{\sin\theta}\delta^{\mu}_3\right)}{(r-i a\cos\theta)\sqrt{2}}  \,. \label{tetrad3}
\end{eqnarray}
Substituting Eq.\ (\ref{tetrad3}) into Eq.\ (\ref{metricstandecom}), we get the rotating metric components $g^{\mu\nu}$.

\subsection{Three classical representations\label{sec2-1}}

\quad In this subsection, we present three classical representations of the rotating black hole,
Eddington-Finkelstein coordinates, Boyer-Lindquist coordinates, and Kerr-Schild coordinates, in order.

The Eddington-Finkelstein form for the geometry is
\begin{eqnarray}
ds^2&=&  - F(r, \theta) dv^2 + 2dv dr -2a\sin^2\theta dr d\psi \nonumber \\
&-&2[1-F(r, \theta)]a\sin^2\theta dv d\psi + \rho^2 d\theta^2 + \frac{\Sigma}{\rho^2} \sin^2\theta d\psi^2 \,, \label{eddfin}
\end{eqnarray}
where $\rho^2=r^2+ a^2\cos^2\theta$, $\Sigma=(r^2+a^2)^2-a^2\triangle\sin^2\theta$, and $\triangle = \rho^2F(r, \theta)+a^2\sin^2\theta$.

To change the metric~\eqref{eddfin} to a Boyer-Lindquist form~\cite{Boyer:1966qh}, we use coordinate transformations
\begin{equation}
dv= dt + \frac{r^2+a^2}{\triangle} dr \,, \quad d\psi=d\phi + \frac{a}{\triangle} dr \,,\label{bolin}
\end{equation}
to obtain
\begin{eqnarray}
ds^2&=&  - F(r, \theta) dt^2 -2[1-F(r, \theta)]a\sin^2\theta dt d\phi + \frac{\Sigma}{\rho^2} \sin^2\theta d\phi^2 + \frac{\rho^2}{\triangle} dr^2 + \rho^2 d\theta^2 \,, \nonumber \\
&=& - \frac{\rho^2\triangle}{\Sigma} dt^2 + \frac{\Sigma}{\rho^2} \sin^2\theta(d\phi-\Omega dt)^2 +  \frac{\rho^2}{\triangle} dr^2 + \rho^2 d\theta^2 \,,
\end{eqnarray}
where $\Omega\equiv - \frac{g_{t\phi}}{g_{\phi\phi}}=\frac{[1-F(r, \theta)]\rho^2 a}{\Sigma}$ measures frame dragging.
Later in this work, we use this coordinates to analyze ergosphere and event horizon.

The Kerr-Schild form~\cite{Kerr:1963ud, KeSch1965} is
\begin{eqnarray}
ds^2&=&  - d \bar{t}^2 +dx^2+dy^2 +dz^2 + \mathcal{F} (dg)^2 \,, \label{kersch}
\end{eqnarray}
where
\begin{equation}
\mathcal{F} = \frac{2Mr^3 -Qr^2 +Kr^{2(2-w)}}{r^4+a^2z^2} \,,
\end{equation}
and
\begin{equation}
dg =d\bar{t}+ \frac{z}{r}dz+ \frac{r(xdx+ydy)+a(ydx-xdy)}{r^2 + a^2} \,.
\end{equation}
Here, we use coordinate transformations,
\begin{eqnarray}
x&=& (r\cos\psi- a\sin\psi)\sin\theta\,, \quad y=(r\sin\psi+a\cos\psi)\sin\theta \,, \nonumber \\
z&=& r\cos\theta \,, \quad \bar{t}=v-r \,.
\end{eqnarray}
The constant $r$ surface is a confocal ellipsoid of rotation about the $z$-axis and is given by
\begin{equation}
\frac{x^2+y^2}{r^2+a^2} + \frac{z^2}{r^2}=1 \,.
\end{equation}
The $(r,z)=0$ region indicates a ring singularity lying on the $xy$-plane, which will be studied in Sec.~\ref{sec3}.

\subsection{The energy-momentum tensor\label{sec2-2}}

\quad The nonvanishing components of the Einstein tensor are given by
\begin{eqnarray}
G_{tt}&=& \frac{2[r^4-2r^3 b + a^2 r^2 -a^4\sin^2\theta\cos^2\theta]b' }{\rho^6} - \frac{r a^2 \sin^2\theta b''}{\rho^4} \,,\nonumber \\
G_{rr}&=&-\frac{2r^2 b'}{\triangle \rho^2}, \qquad G_{\theta\theta} =-\frac{2a^2\cos^2\theta b'}{\rho^2}-r b'' \,, \nonumber \\
G_{t\phi}&=&\frac{2a\sin^2\theta[(r^2+a^2)(a^2\cos^2\theta-r^2)+2r^3 b]b' }{\rho^6} + \frac{r a \sin^2\theta (r^2+a^2) b''}{\rho^4} \,, \nonumber \\
G_{\phi\phi}&=&-\frac{a^2\sin^2\theta[(r^2+a^2)(a^2 +(2r^2+a^2)\cos2\theta)+4r^3\sin^2\theta b]b'}{\rho^6}\nonumber \\
&&-\frac{r\sin^2\theta(r^2+a^2)^2b''}{\rho^4} \,, \label{Etensor}
\end{eqnarray}
where a prime denotes differentiation with respect to $r$ and
\begin{eqnarray}
2b&=&2M -Q^2 r^{-1}+ K r^{1-2w}\,, \quad 2b'= Q^2r^{-2} + (1-2w) K r^{-2w}  \,, \nonumber \\
2b''&=&-2Q^2 r^{-3}-2(1-2w)w K r^{-2w-1} \,.
\end{eqnarray}
When $K=0$, Eq.\ (\ref{Etensor}) reduces to those of Kerr-Newman black hole.
The components $G_{t\phi}$ and $G_{\phi\phi}$ are different from the previous results in \cite{Toshmatov:2015npp, Xu:2016jod}.

Let us consider physical quantities in an orthonormal frame, $(e_{\hat{t}}, e_{\hat{r}}, e_{\hat{\theta}}, e_{\hat{\phi}})$ in Refs.~\cite{Carter:1968ks, Azreg-Ainou:2014nra}, in which the stress-energy tensor for the anisotropic matter field is diagonal,
\begin{eqnarray}
e^{\mu}_{\hat{t}}&=& \frac{(r^2+a^2,0,0,a)}{\rho\sqrt{\triangle}}\,,~~~~e^{\mu}_{\hat{r}}= \frac{\sqrt{\triangle}(0,1,0,0)}{\rho} \,, \nonumber \\
e^{\mu}_{\hat{\theta}}&=&\frac{(0,0,1,0)}{\rho}\,, \quad e^{\mu}_{\hat{\phi}} =\frac{(a\sin^2\theta,0,0,1)}{\rho\sin\theta} \,. \label{otetrad}
\end{eqnarray}

The components of the energy-momentum tensor can be obtained from: $8\pi\varepsilon=e^{\mu}_{\hat{t}}e^{\nu}_{\hat{t}}G_{\mu\nu}$, $8\pi p_{\hat{r}}=e^{\mu}_{\hat{r}}e^{\nu}_{\hat{r}}G_{\mu\nu}$, $8\pi p_{\hat{\theta}}=e^{\mu}_{\hat{\theta}}e^{\nu}_{\hat{\theta}}G_{\mu\nu}$, $8\pi p_{\hat{\phi}}=e^{\mu}_{\hat{\phi}}e^{\nu}_{\hat{\phi}}G_{\mu\nu}$. From (\ref{Etensor}) and (\ref{otetrad}), we get
\begin{eqnarray}
&&\varepsilon=\frac{Q^2 + r^{2w}_o r^{2(1-w)}}{8\pi \rho^{4}}\,,~~
p_{\hat{r}}=-\varepsilon \,, \nonumber \\
&&p_{\hat{\theta}}=p_{\hat{\phi}}
	= [\rho^{2}w-a^2\cos^2\theta] \frac{\varepsilon}{r^2} + (1-w)\frac{Q^2}{8\pi \rho^2 r^2}\,. \label{varepsilon}
\end{eqnarray}
When $a=0$, Eq.~\eqref{varepsilon} reproduces those of the static black hole.

Now, let us examine constraints on $w$ and $K$ from the energy conditions.
In Ref.~\cite{Cho:2017nhx}, the positive energy condition presents $r^{2w}_o=(1-2w)K \geq 0$ for the static black hole.
The condition is extended to $Q^2+ r^{2w}_o r^{2(1-w)} \geq 0$ in the presence of charge.
In general, the positive energy condition presents the same relation as the previous case.
However, for $w \geq 1$, the value $r^{2w}_o$ is allowed to take negative values if one requires the energy condition holds only at the outside of the horizon.
This gives a constraint $r_o^{2w} \geq - Q^2 r_H^{2(w-1)}$, where $r_H$ is the radius of the outmost event horizon.

The flat rotation curve of a galaxy can be explained by the fluid with $w=0$ \cite{Zwicky:1933gu, Rubin:1970zza}.
The dark matter in the galaxy may have energy density decaying like $1/r^2$ outside the core.
However, when $0\leq w \leq 1/2$ the energy density is not localized sufficiently so that the total energy diverges.
For the geometry to be asymptotically flat, therefore, we mainly focus our interest on the system with $w> 1/2$ from now on.
If $w=1$, the matter field describes an extra $U(1)$. The global behaviors of the density and the pressure are the same as the Maxwell field.

\section{Geometric properties\label{sec3}}
\quad In this section, we examine the singularities, the ergosphere and the event horizons.

\subsection{Singularities}
\quad To check the existence of the singularity more clearly, we show the Kretschmann invariant,
\begin{eqnarray}
R^{\alpha\beta\mu\nu}R_{\alpha\beta\mu\nu}&=& R_{KN}+ A K + B K^2 \,,
\end{eqnarray}
where
\begin{eqnarray}
R_{KN}&=& \frac{8}{\rho^{12}}\left[6M^2(r^2-a^2\cos^2\theta)(\rho^4-16^2a^2 \cos^2\theta) -12MQ^2r (r^4-10a^2r^2\cos^2\theta+ 5a^4\cos^4\theta ) \right. \nonumber \\
&+& \left. Q^4(7r^4-34a^2r^2\cos^2\theta + 7a^4 \cos^4\theta) \right] \,,
\end{eqnarray}
\begin{eqnarray}
A&=& \frac{-8r^{-2w}}{\rho^{12}}\left\{ r^6(1+w)[-2Mr(1+2w) + Q^2(1+6w)] +a^2 r^4 [2Mr(31+27w+2w^2)  \right. \nonumber \\
&+& \left. Q^2(-47-31w+10w^2)]\cos^2\theta + a^4r^2 [ Q^2(47-35w+2w^2)+10Mr(-11+3w+2w^2)]\cos^4\theta \right. \nonumber \\
&+& \left. a^6(w-1)[Q^2(1-2w)+6Mr(2w-3)]\cos^6\theta \right\} \,,
\end{eqnarray}
\begin{eqnarray}
B&=& \frac{4r^{-4w}}{\rho^{12}}\left[ r^8(1+5w^2+4 w^3 + 4 w^4) +2a^2r^6(8w^4 -10w^2 -23w-9)\cos^2\theta   \right. \nonumber \\
&+& \left. 2a^4r^4 (12w^4 -12 w^3 -21w^2 -w +29)\cos^4\theta + 2a^6r^2(8w^4 -16w^3-2w^2+19w -9)\cos^6\theta \right. \nonumber \\
&+& \left. a^8(2w^2-3w+1)^2\cos^8\theta \right] \,.
\end{eqnarray}
The $R_{KN}$ part diverges at $\rho = \sqrt{r^2 + a^2 \cos^2\theta} =0$.
Therefore, a ring singularity appears at $r=0$ and $\theta= \pi/2$ when $M\neq 0$.
In addition, for positive $w(\neq 1/2, 1)$, we find that an additional sphere-like singularity exists at $r=0$ because of the $r^{-2w}$ and $r^{-4w}$ proportionalities in $A$ and $B$.

\subsection{Ergosphere and event horizon}

\quad At the static limit surfaces, the timelike Killling vector $\xi^{\mu}_t$ becomes null.
The ergosphere is a region located between the static limit surface and the outer event horizon~\cite{Ruffini:1970sp}, in which both the coordinates $t$ and $r$ are spacelike. Within the ergoregion, a particle cannot remain stationary but can escape to infinity.
\begin{table*}[!ht]
\centering
\begin{tabular}{|c||c|c|} \hline
$ $ & $\mbox{Static limit} $ & ${\rm Event~horizon}$  \\ \hline
$w\neq \frac{1}{2}$ & $\rho^2-2Mr +Q^2 -K r^{2(1-w)}=0$
& $\triangle(r_H) =0$ \\ \hline
$w = \frac{1}{2}$
& $\rho^2+Q^2 -2\bar{M}r\left( 1+ \frac{r_o}{2\bar{M}}\log\frac{r}{r_o} \right)=0$
& $\triangle(r_H)=0$  \\ \hline
\end{tabular}
\caption{\label{tab:1} Static limit and event horizon}
\end{table*}
The event horizon corresponds to a Killing horizon~\cite{Carter:1969zz}, satisfying $\triangle =0$ where
\begin{equation} \label{triangle}
\triangle(r) \equiv
\left\{
\begin{array}{cc}
(r^2+a^2)-2Mr +Q^2 -K r^{2(1-w)} \,, & \quad w \neq 1/2 \vspace{.1cm} \\
(r^2+a^2)+Q^2-2\bar{M}r\left( 1+ \frac{r_o}{2\bar{M}}\log\frac{r}{r_o} \right) \,, & \quad
 w= 1/2
\end{array}
\right. .
\end{equation}
The locations of the event horizon do not depend on the angular coordinates.
The results are summarized in Table \ref{tab:1}, in which $\bar{M}=M+K$.
Hereafter, we focus on the case with $w> 1/2$.
\begin{figure}[t]
\begin{center}
\subfigure[The shape of the function $\triangle(r)$ with $K < 0$. We take $M=0.8$, $a=0.3$ and $Q=0.4$.
The black dashed curve corresponds to a Kerr-Newman black hole, the gray dashed curve to a Kerr one.
The red curve, orange, green, and blue curves correspond to a black hole with $(w, K)=(2/3, -0.1)$, $(2/3, -0.2)$, $(3/2, -0.1)$ and $(3/2, -0.2)$, respectively.]
{\includegraphics[width=3.0in]{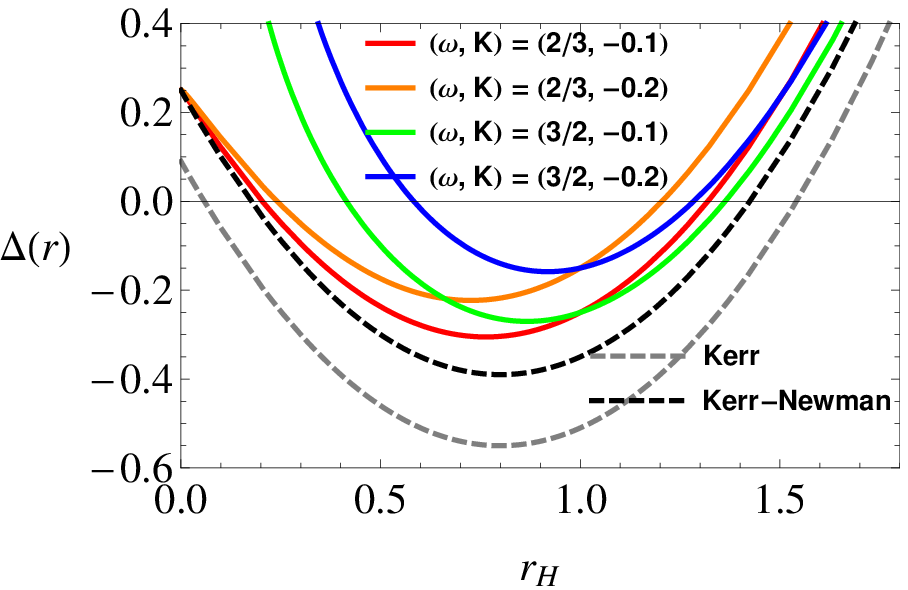}}
~~
\subfigure[The shape of the function $\triangle(r)$ with $K > 0$ and $w=3/2$. We take $M=0.45$ and $a=0.3$. The red, orange, green, and blue curves correspond to a black hole with $(Q, K) =(0.384, 0.015)$, $(0.384, 0.018)$, $(0.424264,0.027)$ and $(0.384, 0.027)$, respectively.]
{\includegraphics[width=3.0in]{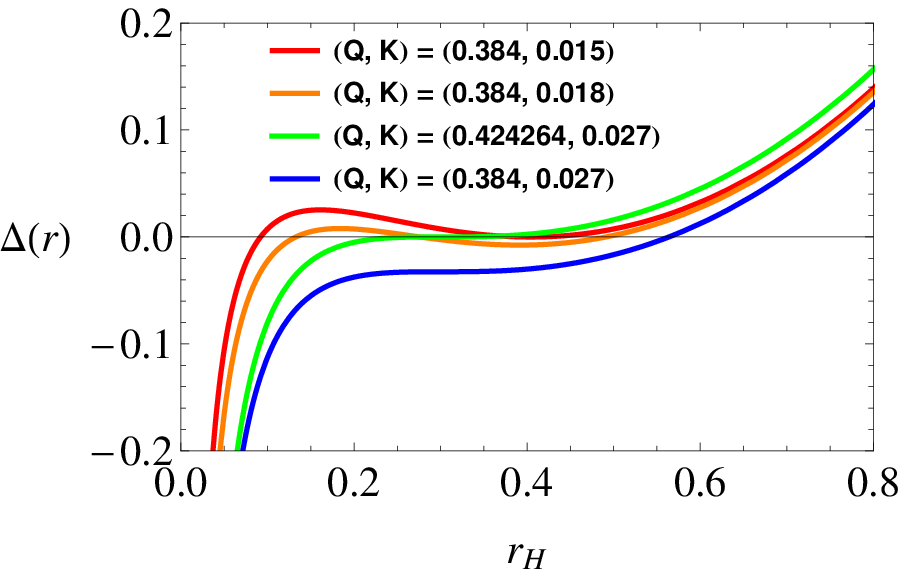}}
\end{center}
\caption{\footnotesize{(color online.)
The shape of the function $\triangle(r)$ in terms of $r$.}}
\label{horizon}
\end{figure}

We presents the shape of $\triangle$ in Fig.~\ref{horizon} for various values of $w$.
The characteristic behavior of $\triangle$ is shown in Fig.~\ref{horizon}(a)  for $K < 0$, where the outer horizon grows as $K$ increases.
Figure~\ref{horizon}(b) shows the characteristic behavior for $K > 0$ and $w > 1$.
The red curve representing an extremal black hole with a doubly-degenerated horizon will go to an extremal Kerr-Newman black hole when $K=0$.
The green curve represents an extremal black hole having a triply degenerated horizon.
This one exists when $K > 0$.

Let us analyze the number of horizons.
The derivatives of $\triangle$ with respect to $r$ are
\begin{eqnarray}
&&\frac{d\triangle}{dr}=\triangle' = 2r-2M-2(1-w) K r^{1-2w} , \nonumber \\
&& \frac{d^2\triangle}{d^2 r}=\triangle'' = 2-2(1-w)(1-2w)K r^{-2w} \,. \label{diffvartri}
\end{eqnarray}

For $K < 0$ and $1/2 < w < 1$, $\triangle$ is dominated by the $r^2$ (constant) term for large (small) $r$.
On the other hand, for $K < 0$ and $w > 1$, $\triangle$ is dominated by the $r^2$ ($K$) term for large (small) $r$.
Therefore, there are two horizons at most in these cases.
Let $\triangle'(r_*)=0$ at $r=r_*$.
Then, it satisfies
\begin{equation}
(1-w)Kr^{1-2w}_*=r_*-M \,.   \nonumber
\end{equation}
We now plug this into Eq.~\eqref{triangle}.
Depending on the value of $\triangle(r_*)$, positive/zero/negative, no/one-degenerate/two horizons exist, respectively.

For $K > 0$ and $w> 1$, $\triangle$ is dominated by the $r^2$ ($K$) term for large (small) $r$.
Because of the linear and the constant terms in $\triangle$, it is possible to have three horizons maximally.
From $\triangle'=0$ we obtain two points, $r_{*+}$ and $r_{*-}$.
If $\triangle'(r_{*+}) \triangle'(r_{*-})>0$, there exists one horizon.
If $\triangle'(r_{*+})\triangle'(r_{*-})=0$, there exist two horizons, one of them is degenerated.
If $\triangle'(r_{*+})\triangle'(r_{*-})<0$, there exist three separated horizons.
We now check the point $r_{**}$ given by $\triangle'(r_{**})=0=\triangle''(r_{**})$.
Then, we obtain relations $r^{2w}_{**}=(1-w)(1-2w)K$ from $\triangle'' =0$, and $r_{**}=\frac{(2w-1)M}{2w}$ from $r_{**}=r_*$.
In this case, the spacetime has the triply-degenerated horizon and its ADM mass becomes $M^2=\frac{4w(w-1)}{(2w-1)^2}(a^2+Q^2)$.

Let us now discuss extremal solutions in detail.
Let $r_{H*}$ represent the location of an outmost degenerated horizon, $\triangle(r_{H*})=0$.
Then, it satisfies
\begin{eqnarray}
(1-w)Kr^{1-2w}_{H*}&=& r_{H*}-M \,,   \label{1st}  \\
\triangle(r_{H*}) &=& \frac{-w}{1-w}\left[ r^2_{H*} - \frac{2w-1}{w} M r_{H*} -\frac{(1-w)(a^2+Q^2)}{w}\right] =0 \,. \label{tri}
\end{eqnarray}
Note that $\triangle(r_{H*}) =0$ presents two real roots, one is positive definite and the other is negative definite when $w< 1$.
For $w\geq 1$, the negative sign fails to represents an outmost horizon.
Therefore, we should choose the positive signature,
\begin{equation} \label{r*}
r_{H*} =\frac{(2w-1)}{2w}M \left[1+ F\right]\,,
\end{equation}
where
\begin{equation}   \label{F}
F \equiv \sqrt{1+ \frac{4w(1-w)}{(2w-1)^2}\frac{(a^2+Q^2)}{ M^2}}\,.
\end{equation}
Then, we can rewrite Eq.~\eqref{1st} to be,
\begin{equation}
(1-w) K r^{2(1-w)}_{H*} = \frac{(2w-1)^2}{4w^2} M^2 (F+1)
\left(F-\frac{1}{2w-1} \right) \,. \label{Feq}
\end{equation}

Let us consider the $1/2 < w<1$ case first.
In Eq.~\eqref{Feq}, the left-hand side(LHS) is negative definite for negative $K$.
This constrains the value of $F$ to be
\begin{equation}  \label{F:con1}
1\leq F\leq \frac{1}{2w-1} \,.
\end{equation}
For $K < 0$ and $w> 1$, the LHS is positive definite in Eq.~\eqref{Feq}.
This constrains the value of $F$ to be
\begin{equation}  \label{F:con2}
\frac{1}{2w-1} \leq F\leq 1 \,.
\end{equation}
For both cases, Eqs.~\eqref{F:con1} and \eqref{F:con2} present a constraint for the extremal solution
\begin{equation}
a^2+Q^2 \leq M^2\,. \nonumber
\end{equation}

For $K > 0$ and $w> 1$, the LHS is negative definite in Eq.~\eqref{Feq}.
This constrains the value of $F$ to be
\begin{equation}  \label{F:con3}
0 \leq F\leq \frac{1}{2w-1}  \,.
\end{equation}
From the definition of $F$ in Eq.~\eqref{F}, we get
\begin{equation}
M^2 \leq a^2+Q^2 \leq \frac{(2w-1)^2}{4w(w-1)} M^2 \,. \nonumber
\end{equation}

As a specific example, let us consider the case with $w= 3/2$.
The locations of the event horizon are obtained from
\begin{equation}
r^2_H-2Mr_H +(a^2+Q^2)- \frac{K}{r_H}=0\,. \label{w=3/2bh}
\end{equation}
The discriminant of this equation is
\begin{equation}
D=-27K^2+4MK(9(a^2+Q^2)-8M^2)-4(a^2+Q^2)^2(a^2+Q^2-M^2) \,. \nonumber
\end{equation}
The roots of $D=0$ gives
\begin{equation}
K=K_\pm\equiv \frac{2}{27}[M(9(a^2+Q^2)-8M^2)\pm\sqrt{(4M^2-3(a^2+Q^2))^3}] \,. \nonumber
\end{equation}
\begin{enumerate}
\item If $K> K_+$ or $K< K_-$ ($D< 0$), there is only one nondegenerate horizon [the blue curve in Fig.~\ref{horizon}(b)].
\item If $K=K_\pm$ ($D=0$), there are two horizons.
One of the two corresponds to a double root at $r_{*\mp}=[2M\mp \sqrt{4M^2-3(a^2+Q^2)}]/3$ [the red curve in Fig.~\ref{horizon}(b)].
If $M=\sqrt{\frac{3}{4}(a^2+Q^2)}$ then $K=\frac{8M^3}{27}$ and the triply degenerated horizon exists at $r_{**}=\frac{2M}{3}$ [the green curve in in Fig.~\ref{horizon}(b)].
\item If $K_- < K < K_+$ ($D>0$), there are three horizons [the orange curve in in Fig.~\ref{horizon}(b)].
\end{enumerate}

\section{Thermodynamics and mass formula \label{sec4}}

\begin{figure}[t]
\begin{center}
\subfigure[Temperature for $K < 0$ and with $a=0.3$ and $Q=0.4$, in which the parameter values are the same as those in Fig.~\ref{horizon}(a)]
{\includegraphics[width=3.0in]{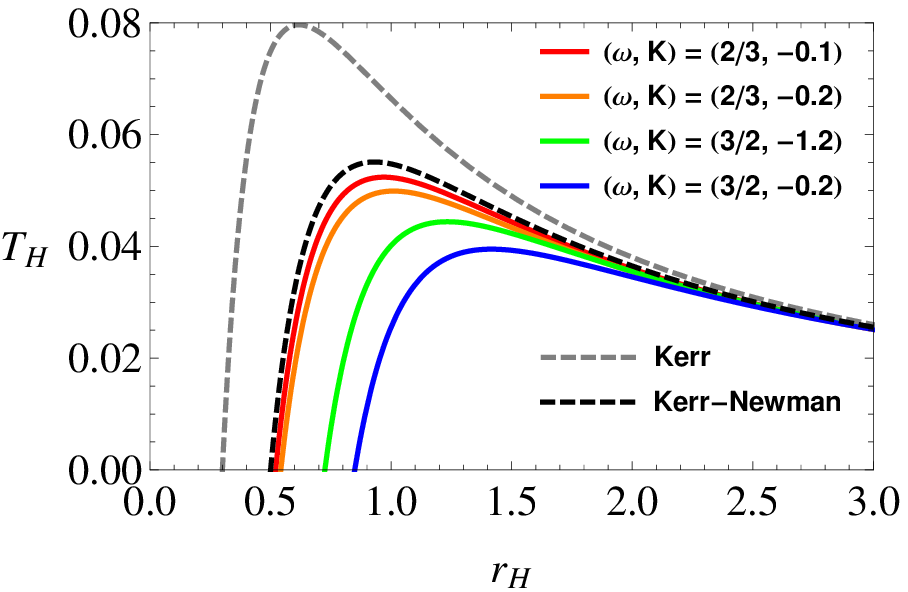}}
~~
\subfigure[Temperature for $K > 0$ and $w=3/2$ and with $a=0.3$, in which the parameter values are the same as those in Fig.~\ref{horizon}(b).]
{\includegraphics[width=3.0in]{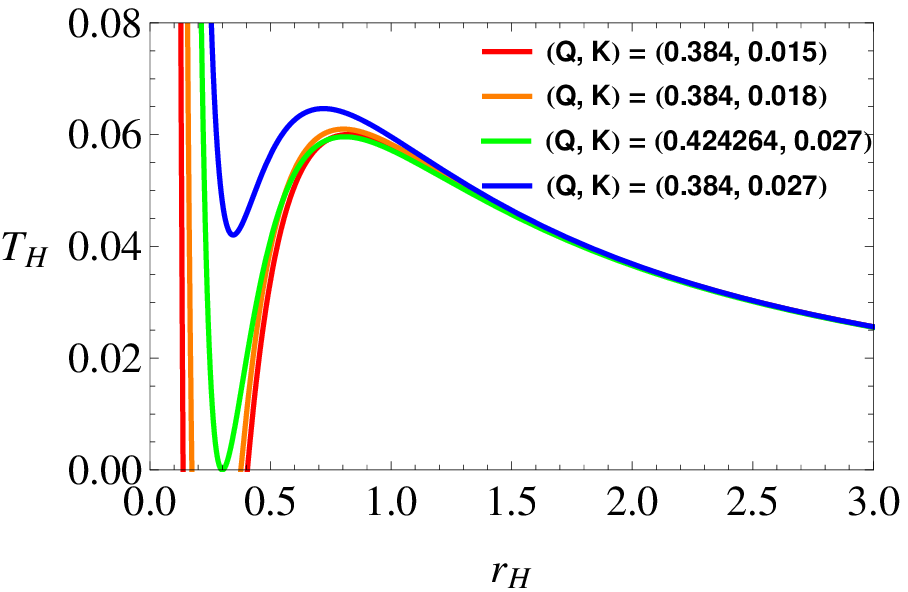}}
\end{center}
\caption{\footnotesize{(color online). Temperature as a function of $r_H$.}}
\label{temp-sheat}
\end{figure}

\quad Let us determine thermodynamic properties of this rotating black hole.
The temperature and the entropy \cite{Bekenstein:1973ur, Hawking:1974sw} are given by
\begin{equation}
T_H=\frac{r_H^2 -(a^2 +Q^2)- (1-2w)K r^{2(1-w)}_H}{4\pi r_H (r^2_H + a^2)},
\qquad  S=\frac{A}{4}=\pi (r^2_H +a^2)   \,. \label{}
\end{equation}
Figure~\ref{temp-sheat} represents the temperature as a function of the horizon radius $r_H$ for the black holes solutions.
Figure~\ref{temp-sheat}(a) and \ref{temp-sheat}(b) show the temperature for the same parameter values as those in Fig.~\ref{horizon}(a) and \ref{horizon}(b), respectively.
We analyze the shape of the temperature curves. We consider only the positive temperature.
Let $r_{H0}$ denote the minimum value of $r_H$ satisfying $T_H(r_{H0}) =0$.
For large $r_H$, $T_H \propto (2\pi r_H)^{-1}$.
On the other hand, for small $r_H$, the $(a^2 +Q^2)$ term dominates the numerator for $1/2 < w < 1$, while the $K r_H^{2-2w}$ term dominates for $w>1$.

When $K<0$, $T_H \to -\infty$ as $r_H\to 0$.
This implies that there exists a value $r_{H0}$ satisfying $T_H(r_{H0}) =0$.
For $r_H \gtrsim r_{H0} $, $T_H$ increases with $r_H$.
Therefore, there exists a maximum value of Hawking temperature at a point $r_{H} (> r_{H0})$, whose location is determined by $T'_H=0$.

When $K>0$ and $w>1$, there exist at least two extremes satisfying $T_H' = 0$.
If the Hawking temperature vanishes at one of the extremes, the geometry describes an extremal black hole with the triply degenerated horizon.

This black hole has nontrivial hairs~\cite{Bardeen:1973gs}.
To express the ADM mass by using Smarr relation \cite{Smarr:1972kt} we consider global charges such as the mass, the charge, the angular momentum and $r_o$, which are measurable by an asymptotic observer.
We rewrite the parameters $K$ with the global charges and $\Phi_o$, the potential field for $r_o$.
The mass formula becomes
\begin{equation}
M= \frac{(M_{KN} + \Phi_o r_o +\Omega_H J)+\sqrt{(M_{KN}+ \Phi_o r_o -\Omega_H J)^2-\frac{4(1-w)}{w}\Phi_o r_o \Omega_H J}}{2}  \,, \label{massfor}
\end{equation}
where $\Omega_H=\frac{a}{r^2_H +a^2}$, $M_{KN}=2 T_H S + 2\Omega_H J +\Phi_H Q$, and $\Phi_H=\frac{r_H Q}{r^2_H+a^2}$ are the angular velocity of the horizon, the mass for a Kerr-Newman black hole and electric potential, respectively,
and
\begin{equation}
\Phi_o=-\frac{Kw r^{3-2w}_H}{r_o(r^2_H+a^2)}=\frac{w}{2w-1} \left( \frac{r_H r_o}{r^2_H+a^2}\right) \left( \frac{r_o}{r_H} \right)^{2(w-1)} \,.
\end{equation}
When $a=0$, this reduces to that of the static black hole~\cite{Cho:2017nhx}.

Let us present the first law of the black hole mechanics.
This law represents a differential relationship between the mass, the entropy, the charge and the angular momentum of the black hole.
For the black hole in the present work, the first law takes the form
\begin{eqnarray}
\delta M=  \frac{M(M-\Omega_H J)} {M(M-\Omega_H J) + \frac{w-1}{w}\Phi_o r_o\Omega_H J} \left[
\delta M_{KN} +\Phi_o \delta r_o + \frac{(w-1)}{w} \frac{\Phi_o r_o \Omega_H }{M-\Omega_H J} \delta J \right] \,, \label{firstlaw}
\end{eqnarray}
where $\delta M_{KN} = T_H \delta S + \Omega_H \delta J +\Phi_H \delta Q$ is that of the Kerr-Newmann black hole.

We now examine the specific heat (heat capacity) to check the local thermodynamic stabilities for rotating black holes.
\begin{figure}[tbh]
\begin{center}
\subfigure[Heat capacity for $K < 0$ and with $a=0.3$ and $Q=0.4$, in which the parameter values are the same as those in Fig.~\ref{horizon}(a)]
{\includegraphics[width=3.0in]{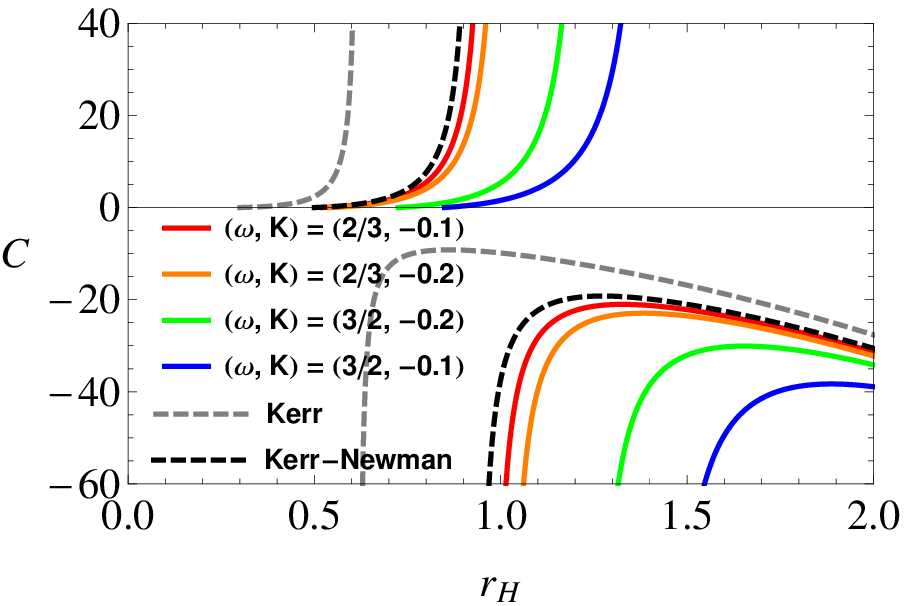}}
~~
\subfigure[Heat capacity for $K > 0$ and $w=3/2$ and with $a=0.3$, in which the parameter values are the same as those in Fig.~\ref{horizon}(b).]
{\includegraphics[width=3.0in]{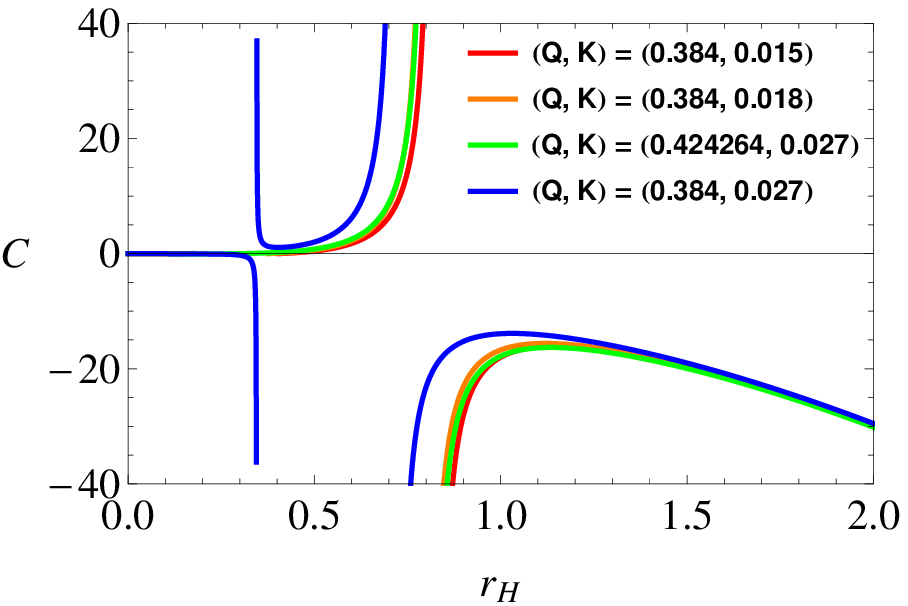}}
\end{center}
\caption{\footnotesize{(color online). Heat capacity as a function of the horizon radius $r_H$.}}
\label{heatcapa}
\end{figure}

The heat capacity, $C=T_H \frac{\partial S}{\partial T_H}$, calculated at constant angular momentum and charge in the canonical ensemble can be obtained as follows:
\begin{eqnarray}
C &=& \frac{2\pi r^2_H (r^2_H + a^2) [r^2_H-(a^2+Q^2)+ (2w-1)Kr^{2(1-w)}_H]}{ a^4 + a^2(Q^2+4r^2_H-(2w-1)^2Kr^{2(1-w)}_H)- r^4_H + 3Q^2r^2_H - (4w^2-1)Kr^{2(2-w)}_H } \,. \label{heatcapacity}
\end{eqnarray}
If $K$ vanishes, this reduces to that of the Kerr-Newman black hole.
\begin{eqnarray}
C &=& \frac{2\pi r^2_H (r^2_H + a^2) [r^2_H-(a^2+Q^2)]}{ a^4 + a^2(Q^2+4r^2_H)- r^4_H + 3Q^2r^2_H  } \,. \nonumber
\end{eqnarray}

Figure \ref{heatcapa} represents the heat capacity as a function of the horizon radius $r_H$ for the black hole solutions.
The black hole is locally stable/unstable if the heat capacity has a positive/negative value, respectively.
In each curve in Fig.~\ref{heatcapa}(a), a singular point is located at the convex up point of the corresponding curve  in Fig.~\ref{temp-sheat}(a), where $|\partial r_H/\partial T_H |\to \infty$.
In Fig.~\ref{heatcapa}(b), two singular points exist in the blue curve.
The right one occurs due to the same reason as Fig.~\ref{heatcapa}(a), while the left one occurs due to the left extremal point of the corresponding curve in Fig.~\ref{temp-sheat}(b).

\section{Energy extraction \label{sec5}}

\quad Let us examine the efficiency of an engine which extracts energy from the rotating black holes.
The irreducible mass represents the minimum mass which cannot be lowered through classical reversible processes.
In principle, we need to obtain the mass energy of a black hole~\cite{Christodoulou:1970wf, Christodoulou:1972kt} by integrating Eq.~\eqref{firstlaw}, which is non-trivial.
Instead, we find the mass energy relation from the horizon information of the black hole:
\begin{equation}
M^2 = \left[M_{I} + \frac{Q^2}{4M_{I}} - \frac{K r^{2(1-w)}_H}{4M_{I}} \right]^2 + \frac{J^2}{4M^2_{I}}  \,, \label{irremass}
\end{equation}
where $M_{I} \equiv \frac{\sqrt{r^2_H +a^2}}{2}$ denotes the irreducible mass of the black hole.
As a special case of the second law of black hole thermodynamics~\cite{Hawking:1971tu}, this is related to the area of the horizon by $A=16\pi M^2_{I}$.
In this sense, the irreducible mass is the final mass of the black hole when its charge or angular momentum is removed by adding external particles through reversible processes which do not modify the black hole entropy.

Through this procedure, $29\%$ and $50\%$ of the mass energy can be extracted for an extremal Kerr black hole and an extremal Kerr-Newman black hole, respectively. It is natural to ask whether or not the present black hole is more efficient than the Kerr-Newman to extract mass energy.

At the present solution, the total mass energy could be decomposed into the rest mass energy, the coulomb energy, the rotational energy, and the energy by the anisotropic matter field.
In our solutions, there exist two types of an extremal black hole. They are extremal black holes with the doubly-degenerated horizon (type I) and the triply degenerated horizon (type II).
Therefore, we can analytically examine both two types of the extremal black hole for $a\rightarrow 0$.
We should employ the more efficient one after comparing both two types. Both two types of solutions are available for $w>1$,
while the type I solution is only available for $1/2 < w< 1$.

Equation~\eqref{irremass} can be simplified
by using the equation for the extreme condition~\eqref{r*}, we get
\begin{equation}
X^2 \equiv \frac{M_I^2}{M^2} = \frac12+\frac{a^2}{4M^2} - G \pm \sqrt{\frac14-G} \,, \label{ratio2}
\end{equation}
where, for $X^2$ to be minimized, one may choose the $-$ sign and
\begin{eqnarray}
G(F) &=& -\frac{(2w-1)^2}{16w^2} (F+1) \left[ F - \frac{2w+1}{2w-1}\right] \,.
\end{eqnarray}
Considering $X^2$ as a function of $G(\leq 1/4)$, it has a global minimum value
$X^2=a^2/4M^2 $ at $G=0$.
It has a local maximum value $X^2(1/4)=1/4+a^2/4M^2$  at $G=1/4$.
The remaining task is to find the physical range of $G$.

Let us consider the case with $1/2<w <1$ first.
In this case, $1\leq F \leq 1/(2w-1)$.
Now, $G(\frac{1}{2w-1}) = (4w)^{-1} > 0$ and $G(1)  = (2w-1)/(4w^2)$.
$G(F)$ takes its maximum at the point $F= F_M$ satisfying
\begin{equation}
G'(F) = -\frac{(2w-1)^2}{8w^2}\left[ F-\frac{1}{2w-1}\right] =0 ; \quad
F= F_M = \frac{1}{2w-1} > 0 \,.
\end{equation}
Because $F_M> 0$, the range of $G$ becomes $0< G(1) \leq G \leq G(F_M)$.
Therefore, the minimum value of $X^2$  will be given at $G=G(1)$,
\begin{equation}
X^2 = \frac{a^2}{4M^2} + \frac{(2w-1)^2}{4w^2}, \qquad  \frac{1}{2}< w< 1 \,.
\end{equation}

Next, we consider the case with $w\geq 1$.
In this case, the range of $F$ is $0\leq F \leq 1$.
Then, $G(0) = (4w^2-1)/(16w^2)>0 $ and $G(1) = (2w-1)/(4w^2)>0$.
In this range of $F$, $G(F)$ has its maximum $G(F_M) $ at $F=F_M$.
Let us compare $G(1)$ and $G(0)$,
\begin{equation}
G(0)-G(1) = \frac{(2w-1)(2w-3)}{16w^2} \,.
\end{equation}
For $ 1\leq w\leq 3/2$, $G(0) \leq G(1)$.
Therefore, the range of $G$ becomes $G(0)$ to $1/4$.
The minimum value will be given by
\begin{equation}
X^2(G(0)) = \frac{a^2}{4M^2}+ \frac{(2w-1)^2}{16w^2} \,, \qquad 1 \leq w\leq \frac{3}2 \,.
\end{equation}
For $w\geq 3/2$, $G(1) \leq G(0)$.
Therefore, the minimum value becomes
\begin{equation}
X^2(G(1)) = \frac{a^2}{4M^2} + \frac{1}{4w^2} \,, \qquad w\geq \frac32 \,.
\end{equation}
Now, the ratio of the extracted mass energy relative to the ADM mass is given by
\begin{equation}
\frac{\Delta M}{M} \equiv 1-\frac{M_I}{M}= 1-X \,.
\end{equation}

\begin{figure}[H]
\begin{center}
{\includegraphics[width=4.0in]{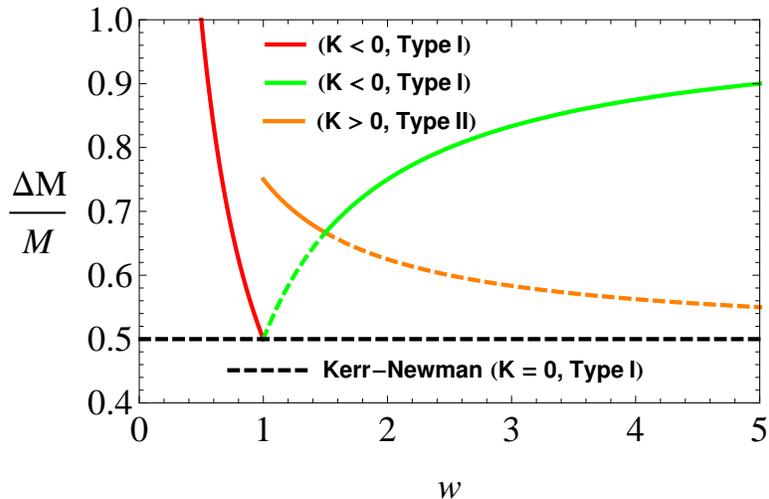}}
\end{center}
\caption{\footnotesize{(color online).
The best efficiency of mass extraction with respect to the equation of state $w$.
}}
\label{extractenergy}
\end{figure}
Figure \ref{extractenergy} represents the extracted mass energy as a function of $w$ for the extremal black hole solutions.
The black dashed curve represents the efficiency for the Kerr-Newman black hole ($50\%$).
Solid curves represent the best efficiency for the present black holes having anisotropic matter which is achieved with the $a\rightarrow 0$ limit. The red and the green curves represent those with $K < 0$ and for type I solutions, while the orange curve represents that $K > 0$ and for type II solutions, respectively. The dashed green curve represents the best efficiency when the positive energy condition is imposed inside the horizon.

\section{Summary and discussions \label{sec6}}

\quad We have presented a family of new rotating black hole solutions to Einstein's equations with the anisotropic matter field.
The rotating geometry was obtained from the known static solution by employing the Newman-Janis algorithm.

We chose an orthonormal tetrad as a proper reference frame for measuring the energy density and the pressure.
On that frame, the energy-momentum tensor becomes diagonal and takes a simple form.
Our solutions correspond to a four parameter family of axisymmetric stationary spacetimes.
The new term with $K$ decays more rapidly than the term with a mass and more slowly than the term with a electric charge for $1/2 < w < 1$, while the term decays more rapidly than the term with the charge for $w > 1$.
After introducing a potential for the anisotropic matter, we have obtained the mass formula, the Smarr relation.
The potential was written by means of a new parameter, which is related to a conserved charge for the anisotropic matter.
The first law of thermodynamics was also obtained.

We have considered a set of successive reversible processes extracting energy from the black hole and calculated the efficiency.
For a Kerr-Newman black hole, the most efficient engine corresponds to that from an extremely charged black hole to a neutral one~\cite{Christodoulou:1972kt}.

For the present solutions having anisotropic matter, we have  found that the efficiency is better than that of the Kerr-Newman.
If we require the energy condition be satisfied over the whole spacetime, the best efficiency is achieved for the extremal black hole having a doubly-degenerated horizon.
On the other hand, the energy condition is loosend to be satisfied only outside of the event horizon, a higher efficiency is achieved for the extremal black hole having the triply degenerated horizon when $1\leq w\leq 3/2$, where $w$ denotes the equation of state.

Our black hole solutions avoid the no-hair conjecture(theorem)\cite{Herdeiro:2014goa, Ruffini:1971bza, Antoniou:2017acq, Lee:2018zym} in Einstein-Maxwell theory because a nontrivial matter field exists outside the horizon.
This is possible thanks to the negative radial pressure similar to an electrostatic field.
When $w=1$, the new potential has a new $U(1)$ charge.

The physical reality of the solutions will be tested in future observations.
The gravitational(dynamical) and the global thermal stabilities of the solutions should be tested and the quasinormal modes should be found.
Fortunately, the static solutions with an anisotropic matter field are stable under radial perturbations~\cite{Cho:2017nhx}.
The global thermodynamic stability can be determined by comparing the free energy  between the black hole and the reference background \cite{Hawking:1982dh, Cai:2001dz, Khimphun:2016gsn}.
The Gibbs free energy in the grand canonical ensemble corresponding to a fixed electric potential and temperature takes the form
\begin{equation*}
G= M - T_HS -\Omega_H J - \Phi_H Q- \Phi_o r_o - \frac{(w-1)}{w}\frac{\Phi_o r_o \Omega_H J}{(M-\Omega_H J)}  \,, \label{}
\end{equation*}
where $M$ is interpreted as the enthalpy.
The phase transition occurs when the two free energies are the same.
Such issues will be treated for future works.

\section*{Acknowledgments}
H.-C.~K. and Y.~L. (NRF-2020R1A2C1009313), B.-H.~L.
(NRF-2019R1F1A1062730), and W.~L. (NRF-2016R1D1
A1B01010234) were supported by Basic Science Research
Program through the National Research Foundation of
Korea funded by the Ministry of Education. We would
like to thank Hyun Kyu Lee for helpful discussions,
Gungwon Kang and Hong Lu for helpful comments at
STGCOS(String theory, Gravitation, and Cosmology) in Pohang, and Dong-han Yeom for crosschecking
the Einstein tensor using MAPLE program. We
would like to thank Kimyeong Lee and Miok Park for
their hospitality during our visit to KIAS, Seoktae Koh to
Jeju National University, Yun Soo Myung to Inje
University. W. Lee also would like to thank Hongsu
Kim for introducing the Newman-Janis algorithm and
Hocheol Lee for helping the numerical work.

\end{document}